\documentstyle[12pt]{article}
\begin{document}
\baselineskip 22pt
\pagestyle{plain}

\rightline{JHU-TIPAC-930011 // DFTT 22/93}
\vskip0.5truecm
\centerline{\Large\bf On the Treatment of Neutrino Oscillations }
\centerline{\Large\bf Without Resort to Weak Eigenstates}
\vskip0.5truecm
\centerline{\large C.~Giunti}
\centerline{\it INFN, Sezione di Torino }
\vskip -0.25truecm
\centerline{\it I--10125 Torino, Italy }
\vskip0.4truecm
\centerline{\large C.W.~Kim and J.A.~Lee }
\centerline{\it Department of Physics and Astronomy }
\vskip -0.25truecm
\centerline{\it The Johns Hopkins University }
\vskip -0.25truecm
\centerline{\it Baltimore, Maryland 21218, USA }
\vskip0.4truecm
\centerline{\large U.W.~Lee }
\centerline{\it Department of Physics }
\vskip -0.25truecm
\centerline{\it Mokpo University }
\vskip -0.25truecm
\centerline{\it Mokpo,Korea }
\vspace*{0.4truecm}
\centerline{\large Abstract }
\vspace*{0.3truecm}
We discuss neutrino oscillations in the framework of
the quantum field theory
without introducing the concept of neutrino weak eigenstates.
The external particles are described by wave packets
and the different mass eigenstate neutrinos
propagate between the production and detection interactions,
which are macroscopically localized in space-time.
The time-averaged cross section,
which is the measurable quantity
in the usual experimental setting,
is calculated.
It is shown that
only in the extremely relativistic limit
the usual quantum mechanical oscillation probability
can be factored out of the cross section.

\noindent \underline{$\phantom{space space space}$}

\noindent
{\scriptsize
GIUNTI@TORINO.INFN.IT;~CWKIM@JHUVMS.HCF.JHU.EDU;~LEEJ@DIRAC.PHA.JHU.EDU}
\newpage

\vspace*{1.5cm}
\section{Introduction}

Neutrino oscillations have long been recognized
as a powerful tool to probe the intrinsic properties of neutrinos
\cite{BilenkyPontecorvo78}.
Furthermore, it already appears
that they may provide an elegant solution to the Solar Neutrino Problem,
possibly leading to information on
the basic properties of neutrinos such as mass and mixing angle.

If neutrinos are massive and mixed,
a weak eigenstate neutrino
which is produced
in a weak process accompanying a lepton is
a linear superposition of mass eigenstates.
In the standard treatment of neutrino oscillations
\cite{BilenkyPontecorvo78,BilenkyPetcov87},
the mass eigenstates are assumed to be relativistic
and to have the same momentum but different energies.
Because of the energy differences,
the quantum mechanical probability
of finding weak eigenstates becomes a function of the distance
from the production point, leading to neutrino oscillations.

Although the standard approach of treating neutrino oscillations with use of
the weak eigenstates is physically intuitive and simple,
it is, strictly speaking, neither rigorous  nor  sufficient for a {\em
complete}
understanding of the physics involved in neutrino oscillations.
Furthermore, as shown in Ref.\cite{GKL9119},
the usual ``weak eigenstates''
$ \left| \nu_{\alpha} \right\rangle =
  \sum_{a}\ {\cal U}_{\alpha a}^{*} \left| \nu_{a} \right\rangle $
($ {\cal U} $ is the mixing matrix of the neutrino fields
and $ \left| \nu_{a} \right\rangle $ are the mass eigenstates)
describe correctly the neutrinos produced and detected
in weak-interaction processes
{\em only in the extremely relativistic limit}.
Also,
energy-momentum conservation in the process
in which the neutrino is created implies that
the different mass-eigenstate components must have different momenta as well as
different energies
\cite{Winter81}.
On the other hand,
if the particles involved in the production
(as well as the detection) process
are assumed to have definite four-momenta,
then the neutrino produced (detected) is forced to have
a definite four-momentum,
implying that the neutrino is one of
the mass eigenstates \cite{Shrock80}.
This observation suggests an apparent
incompatibility between
energy-momentum conservation
in the production and detection processes
and the neutrino oscillations.
In other words, energy-momentum conservation which forces neutrinos to be in
mass eigenstates
is incompatible with neutrino oscillations.
This apparent incompatibility, however, does not arise in a physical situation
since a necessary condition for neutrino oscillations to occur is that
the neutrino source and detector
are localized within a region much smaller than the
oscillation length,
and hence the neutrino momentum must have
at least the corresponding spread given by the
uncertainty principle
\cite{Kayser81}.
This spread is responsible for neutrino oscillations.

The localization of the neutrino source and detector
and the spread of the neutrino momentum
imply that a propagating flavor neutrino
is not described by a superposition of plane waves,
but instead by a superposition of localized wave packets
\cite{Kayser81}.
The wave packet treatment necessary
for a correct quantum mechanical description of neutrino oscillations
has been discussed in Ref.\cite{GKL9116}.
However,
neutrino oscillations
have, so far, been discussed in the framework of quantum mechanics of
neutrino propagation,
whereas the effects of the production and detection weak-interaction processes
have been neglected.
As shown in Ref.\cite{GKL9119},
the neutrino oscillation probability
is independent from the details
of the production and detection processes
{\em only in the extremely relativistic limit}.
Hence a quantum field theoretical treatment
of neutrino oscillations
is necessary for the discussion
of the case in which some of the mass eigenstates
happen to be not extremely relativistic.
It is, of course, expected that
a quantum field theoretical treatment
must reproduce the quantum mechanical oscillation probability
in the extremely relativistic limit.

In this paper, we present the quantum field theoretical treatment
of neutrino oscillations by using, as an example,
a specific flavor changing process (see Eq.(\ref{E:0}))
and by calculating its cross section.
The external (initial and final) particles
are described by wave packets
and the mass eigenstate neutrinos
propagate from the production region to a detector
which are macroscopically separated in space-time.
Since energies and momenta of the external particles
are not precisely defined and
energy-momentum is conserved in the interaction vertices
within the  uncertainty principle,
the contributions from the propagation of different mass eigenstate neutrinos
can interfere to produce oscillations.

In the quantum field theoretical treatment
it is impossible to derive a {\em general formula} for neutrino oscillations
because the cross sections depend on the details of the specific
production and detection interactions involved.
Hence in this paper we illustrate a general method using a specific example.
Furthermore, strictly speaking, an oscillation probability
cannot even be defined because the space dependence of the cross section
cannot be factorized out.
This does not imply that neutrino oscillation phenomena do not take place.
The phenomena can only be inferred from actual measurements of cross sections.
It will be shown that
the quantum field theoretical treatment yields
the standard quantum mechanical oscillation probability
{\em only in the extremely relativistic limit}.

The plan of this paper is as follows:
In Section \ref{S:A}
we present the calculation of the amplitude
for a specific flavor changing process.
In Section \ref{S:CS}
we calculate the cross section
and its time average,
which is an experimentally measurable quantity.

Finally, we discuss, in Section \ref{S:RL},
the case of extremely relativistic neutrinos and reproduce the standard
oscillation probability.

\vspace*{1.5cm}
\section{ Amplitude }
\label{S:A}

Let us consider the weak flavor-changing processes
\begin{equation} \vphantom{\Bigg|}
\begin{array}{c}
P_I \to P_F + \mu^{+} + \nu
\hskip5pt
\hphantom{ \searrow }
\hskip5pt
\hphantom{ \nu + D_I \to D_F + e^{-} }
\\[-10pt]
\hphantom{ P_{I} \to P_{F} + \mu^{+} + \nu }
\hskip5pt
\searrow
\hskip5pt
\hphantom{ \nu + D_{I} \to D_{F} + e^{-} }
\\[-10pt]
\hphantom{ P_{I} \to P_{F} + \mu^{+} + \nu }
\hskip5pt
\hphantom{ \searrow }
\hskip5pt
\nu + D_I \to D_F + e^{-}\ ,
\end{array}
\label{E:0}
\end{equation}
occurring through the intermediate propagation of a neutrino,
where $ P_{I} $ and $ P_{F} $ ($ D_{I} $ and $D_{F} $)
are the initial and final
production (detection)
particles.
For simplicity, we denote the energy-momentum four vectors for the particles
involved in Eq.(\ref{E:0}) as
\begin{equation}
p=p^\prime+p_\mu+p_\nu\phantom{abc}, \phantom{abc} p_\nu+k=k^\prime+p_e\ .
\label{E:01}
\end{equation}
We will consider a process in which the production and detection interactions
are macroscopically localized at the coordinates
$ ( \vec{X}_{P} , T_{P} ) $ and $ ( \vec{X}_{D} , T_{D} ) $,
respectively.

The relevant weak interaction Lagrangians are
\begin{eqnarray} &\vphantom{\Bigg|}&
{\cal L}_{P}(x) = \frac{G_F}{\sqrt{2}}
\sum_{a} {\cal U}_{\mu a}^*
\overline{\nu_{a}}(x) \gamma^\alpha (1+\gamma_5) \mu(x) J^P_\alpha(x)
\nonumber \\[-2.5mm]
\label{E:1} \\[-2.5mm] &\vphantom{\Bigg|}&
{\cal L}_{D}(x) = \frac{G_F}{\sqrt{2}}
\sum_{a} {\cal U}_{ea}
\overline{e}(x) \gamma^\alpha(1+\gamma_5) \nu_{a}(x) J^{D}_\alpha(x)\ ,
\nonumber
\end{eqnarray}
where
$ J^P_\alpha(x) $ and $ J^D_\alpha(x) $
are the weak currents of the production and detection particles, respectively
and the other notations are self-evident.

The amplitude for the process is
\begin{eqnarray} &\vphantom{\Bigg|}&
{\cal A} =
\left\langle
{ P_{F} , \mu^{+} , D_{F} , e^{-} }
\vphantom{
{T \left[ \int d^4x_1 \, d^4x_2 \, {\cal L}_{P}(x_1) {\cal L}_{D}(x_2) \right]
}
{ P_{I} , D_{I} }}
\right|
\vphantom{ { P_{F} , \mu^{+} , D_{F} , e^{-} }}
{T \left[ \int d^4x_1 \, d^4x_2 \, {\cal L}_{P}(x_1) {\cal L}_{D}(x_2) \right]
}
\vphantom{ { P_{I} , D_{I} }}
\left|
\vphantom{
{ P_{F} , \mu^{+} , D_{F} , e^{-} }
{T \left[ \int d^4x_1\, d^4x_2 \, {\cal L}_{P}(x_1) {\cal L}_{D}(x_2) \right]
}}
{ P_{I} , D_{I} }
\right\rangle
\ ,
\label{E:2}
\end{eqnarray}
where the initial and final particles are described by the wave packets
\begin{eqnarray} &\vphantom{\Bigg|}&
\left |  P_{I} \right\rangle = \int { d\vec{p} \over (2\pi)^{3/2} }
\ \psi_{P_{I}}(\vec{p};\vec{X}_{P},T_{P},\left\langle{\vec{p}}\right\rangle   )
\ \left|  P_{I}(\vec{p}) \right\rangle
\nonumber \\ &\vphantom{\Bigg|}& \phantom{ } \hskip 2truecm \vdots
\label{E:3} \\ &\vphantom{\Bigg|}& \phantom{ }
\left| e^{-} \right\rangle  = \int { d\vec{p}_{e} \over (2\pi)^{3/2} }
\
\psi_{e}(\vec{p}_{e};\vec{X}_{D},T_{D},\left\langle{\vec{p}_{e}}\right\rangle)
\ \left|  e^{-}(\vec{p}_{e}) \right\rangle.
\nonumber
\end{eqnarray}
The form of the wavefunctions $ \psi $ in momentum space
is determined by how the initial particles are prepared
and how the final particles are detected.
In the following we will assume, for simplicity,
gaussian wavefunctions,
whose form is given by Eq.(\ref{E:A.1}) in Appendix A.
In Eq.(\ref{E:3}), $\left\langle{\vec{p}}\right\rangle,\cdots,
\left\langle{\vec{p}_e}\right\rangle$
are the average momenta of the particles around which their
momenta are spread due to the uncertainty principle.
The wave packets are constructed in such a way that
at the time $ t = T_{P} $
the wave packets of the muon and the production particles
overlap at $ \vec{x} \simeq \vec{X}_{P} $ and
at the time $ t = T_{D} $
the wave packets of the electron and the detection particles
overlap at $ \vec{x} \simeq \vec{X}_{D} $.

The propagators of the mass eigenstate neutrinos are
\begin{equation} \vphantom{\Bigg|}
\left\langle{0}
\vphantom{{ T \left[ \nu_{a}(x_2) \overline{\nu_{a}}(x_1) \right] }{0}}\right|
\vphantom{{0}}{ T \left[ \nu_{a}(x_2) \overline{\nu_{a}}(x_1) \right] }
\vphantom{{0}}
\left|\vphantom{{0}{ T \left[ \nu_{a}(x_2) \overline{\nu_{a}}(x_1) \right] }}
{0}\right\rangle
= i \int { d^4q \over (2\pi)^4 }
\ \frac{ q\kern-.46em/ + m_{a} }{ q^2 - m_{a}^2 + i \epsilon }
\ {\rm e}^{ - i q \cdot \left( x_2 - x_1 \right) }\ .
\label{E:4}
\end{equation}
Hence the amplitude in Eq.(\ref{E:2}) becomes, with Eqs. (\ref{E:3})
and (\ref{E:4}),
\begin{eqnarray} &\vphantom{\Bigg|}&
{\cal A} =
i\ G_F^2
\ \sum_{a}\ {\cal U}_{\mu a}^{*}\ {\cal U}_{ea}
\int d^4x_1 \int d^4x_2
\int { d^4q \over (2\pi)^4 }\ {\rm e}^{ - i q \cdot \left( x_2 - x_1 \right) }
\nonumber \\ &\vphantom{\Bigg|}& \phantom{ {\cal A} = } \times
\int { d\vec{p} \over (2\pi)^{3/2} }\
	\psi_{P_{I}}\ {\rm e}^{ - i p \cdot x_1 }
\int { d\vec{k} \over (2\pi)^{3/2} }\
	\psi_{D_{I}}\ {\rm e}^{ - i k \cdot x_2 }
\nonumber \\ &\vphantom{\Bigg|}& \phantom{ {\cal A} = } \times
\int { d\vec{p}^{\phantom{i}\prime} \over (2\pi)^{3/2} }\
	\psi_{P_{F}}^{*}\ {\rm e}^{ i p^\prime \cdot x_1 }
\int { d\vec{p}_{\mu} \over (2\pi)^{3/2} }\
	\psi_{\mu}^{*}\ {\rm e}^{ i p_{\mu} \cdot x_1 }
\label{E:5} \\ &\vphantom{\Bigg|}& \phantom{ {\cal A} = } \times
\int { d\vec{k}^\prime \over (2\pi)^{3/2} }\
	\psi_{D_{F}}^{*}\ {\rm e}^{ i k^\prime \cdot x_2 }
\int { d\vec{p}_{e} \over (2\pi)^{3/2} }\
	\psi_{e}^{*}\ {\rm e}^{ i p_{e} \cdot x_2 }
\nonumber \\ &\vphantom{\Bigg|}& \phantom{ {\cal A} = } \times
J^{D}_{\lambda}(k,k^\prime)\
\overline{u_{e}}(p_{e})\
\gamma^\lambda
\frac{ (1+\gamma_5) q\kern-.46em/ }{ q^2 - m_{a}^2 + i \epsilon }\
\gamma^\varrho\
v_{\mu}(p_{\mu})\
J^{P}_{\varrho}(p,p^\prime)\ ,
\nonumber
\end{eqnarray}
where  $J^{D}_{\lambda}(k,k^\prime)$ and $J^{P}_{\varrho}(p,p^\prime)$
are the matrix elements of the weak currents of the detection and
production particles, respectively
and we have not explicitly written down the arguments of the wave packets.

The momentum integrations of external particles can easily be carried out if
the wave packets in momentum space are sharply peaked around
their average momenta (we have assumed, for simplicity,
that all the wave packets are gaussian with the same width $\sigma_{x}$).
After straightforward but tedious integrations,
the amplitude can be written, with appropriate changes of the coordinates,
as
\begin{equation}
\begin{array}{rcl} \displaystyle
{\cal A}
& \propto & \displaystyle
\sum_{a} {\cal U}_{\mu a}^{*} \, {\cal U}_{ea}
\int { d^4q \over (2\pi)^4 } \,
\overline{U_{D}} \,
\frac{ q\kern-.46em/ }{ q^2 - m_{a}^2 + i \epsilon } \,
V_{P} \,
\exp\left[ - i q_0 T + i \vec{q} \cdot \vec{L} \right]
\\ \displaystyle
& & \displaystyle
\times
\int d^4x_1 \,
\exp\left[
- i \left( E_P - q_0  \right) t_1
+ i \left( \vec{p}_P - \vec{q} \right) \cdot \vec{x}_1
- 3 \, {\scriptstyle{ \vec{x}_1^2 }\over
         \scriptstyle { 4 \sigma_{x}^2 }}
+ 3 \, {\scriptstyle{ \vec{v}_{P} \cdot \vec{x}_1 }
           \over\scriptstyle  { 2 \sigma_{x}^2 } }\, t_1
-{ \scriptstyle{ \left( \vec{v}^2
	       + \vec{v}^{\prime 2}
               + \vec{v}_{\mu}^2 \right) }\over
   \scriptstyle { 4 \sigma_{x}^2 }} \, t_1^2
    \right]
\\ \displaystyle
& & \displaystyle
\times
\int d^4x_2 \,
\exp\left[
- i \left( E_D + q_0 \right) t_2
+ i \left( \vec{p}_D + \vec{q} \right) \cdot \vec{x}_2
- 3 \, {\scriptstyle { \vec{x}_2^2 }\over
        \scriptstyle { 4 \sigma_{x}^2 }}
+ 3 \, {\scriptstyle { \vec{v}_{D} \cdot \vec{x}_2 }\over
        \scriptstyle { 2 \sigma_{x}^2 }} \, t_2
- 	{{\scriptstyle  \left( \vec{u}^2
               + \vec{u}^{\prime 2}
               + \vec{v}_{e}^2 \right)} \over
    	 {\scriptstyle  4 \sigma_{x}^2 }} \, t_2^2
    \right]\ ,
\end{array}
\label{E:8}
\end{equation}
where we have defined
\begin{equation}
\begin{array}{lcl} \displaystyle \vphantom{Bigg|}
E_P \equiv \left\langle{p_0}\right\rangle
	- \left\langle{p_0^\prime}\right\rangle
	- \left\langle{E_{\mu}}\right\rangle
& & \displaystyle
E_D \equiv \left\langle{k_0}\right\rangle
	- \left\langle{k_0^\prime}\right\rangle
	- \left\langle{E_{e}}\right\rangle
\\ \displaystyle \vphantom{Bigg|}
\vec{p}_P \equiv \left\langle{\vec{p}}\right\rangle
	- \left\langle{\vec{p}^\prime} \right\rangle
- \left\langle{\vec{p}_{\mu}}\right\rangle
& & \displaystyle
\vec{p}_D \equiv \left\langle{\vec{k}}\right\rangle
	- \left\langle{\vec{k}^\prime}\right\rangle
- \left\langle{\vec{p}_{e}}\right\rangle
\\ \displaystyle \vphantom{Bigg|}
3 \vec{v}_{P} \equiv \vec{v} + \vec{v}^\prime + \vec{v}_{\mu}
& & \displaystyle
3 \vec{v}_{D} \equiv \vec{u} + \vec{u}^\prime + \vec{v}_{e}
\\ \displaystyle \vphantom{Bigg|}
T \equiv T_D-T_P
& & \displaystyle
\vec{L} \equiv \vec{X}_D-\vec{X}_P
\\ \displaystyle \vphantom{Bigg|}
\overline{U_{D}} \equiv
J^{D}_{\alpha}
	\left(\left\langle{k}\right\rangle,
		\left\langle{k^\prime}\right\rangle\right)
\overline{u_{e}}
	\left(\left\langle{p_{e}}\right\rangle\right)
\gamma^\alpha(1+\gamma_5)
& & \displaystyle
V_{P} \equiv
(1+\gamma_5) \gamma^\alpha
v_{\mu}\left(\left\langle{p_{\mu}}\right\rangle\right)
J^{P}_\alpha \left(\left\langle{p}\right\rangle,
		\left\langle{p^\prime}\right\rangle\right)\ .
\end{array}
\label{E:81}
\end{equation}
Also in Eqs.(\ref{E:8}) and (\ref{E:81}), we have introduced the notation
\begin{displaymath}
\begin{array}{lclclclclcl}
	\vec{v}&=&\displaystyle
		\frac{\langle\vec{p}\rangle}{\langle p_0\rangle}&\; ,\;&
	\vec{v}^\prime&=&\displaystyle
		\frac{\langle\vec{p^\prime}\rangle}
		{\langle p_0^\prime\rangle}&\; ,\;&
	\vec{v}_\mu &=&\displaystyle
		\frac{\langle\vec{p_\mu}\rangle}
		{\langle E_\mu \rangle}\\
	\vec{u}&=&\displaystyle
		\frac{\langle \vec{k}\rangle }
		{\langle k_0\rangle }&\; ,\;&
	\vec{u}^\prime&=&\displaystyle
		\frac{\langle \vec{k}^\prime\rangle }
		{\langle {k_0^\prime}\rangle }&\; ,\;&
	\vec{v}_e &=&\displaystyle
		\frac{\langle\vec{p_e}\rangle}
		{\langle E_e \rangle}\ .
\end{array}
\end{displaymath}
Carrying out the integrals over $x_1$ and $x_2$ which are gaussian,
we obtain, from Eq.(\ref{E:8}),
\begin{equation}
\begin{array}{rcl} \displaystyle
{\cal A}
& \propto & \displaystyle
\sum_{a} {\cal U}_{\mu a}^{*} \, {\cal U}_{ea}
\int { d^4q \over (2\pi)^4 } \,
\overline{U_{D}} \,
\frac{ q\kern-.46em/ }{ q^2 - m_{a}^2 + i \epsilon } \,
V_{P} \,
\exp\left[ - i q_0 T + i \vec{q} \cdot \vec{L} \right]
\\ \displaystyle
& & \displaystyle
\times
\exp\left[
- \frac{ \left( \vec{p}_P - \vec{q} \right)^2 }
       {12 \sigma_{p}^2 }
- \frac{ \left[ \left( E_P - q_0  \right)
              - \left( \vec{p}_P - \vec{q} \right)
	        \cdot \vec{v}_{P} \right]^2 }
       {12 \sigma_{p}^2 \lambda_P }
\right.
\\ \displaystyle
& & \displaystyle
\phantom{ \times \exp[ }
\left.
- \frac{ \left( \vec{p}_D + \vec{q} \right)^2 }
       { 12 \sigma_{p}^2 }
- \frac{ \left[ \left( E_D + q_0 \right)
	      - \left( \vec{p}_D + \vec{q} \right)
	        \cdot \vec{v}_{D} \right]^2 }
       { 12 \sigma_{p}^2 \lambda_D }
\right]
\end{array}
\label{E:11}
\end{equation}
where
\begin{equation}
\lambda_P  \equiv  \displaystyle \frac{1}{3}\left(
	\vec{v}^2+\vec{v}^{\prime 2}+\vec{v}_{\mu}^2 \right)
	-\vec{v}_P^2
\phantom{space}
\lambda_D  \equiv  \displaystyle \frac{1}{3}\left(
	\vec{u}^2+\vec{u}^{\prime 2}+\vec{v}_e^2 \right)
	-\vec{v}_D^2\ .
\label{E:10a}
\end{equation}

We now face the problem
of performing the integration over $q$.
In usual calculations of the processes
occurring through the propagation
of a virtual intermediate particle
the integration over its four-momentum $q$
is easily simplified
by the Dirac $\delta$-functions
arising from energy-momentum conservation
in the interaction vertices.
On the other hand,
in the wave packet treatment
of the initial and final particles,
energy-momentum is not exactly conserved
and there are no Dirac $\delta$-functions
available for the simplification of the integration over $q$.
However,
in our case
the production and detection interactions
are macroscopically separated,
so only the propagation of real neutrinos
contribute significantly to the process.
This physical fact
allows us to perform the integration over $q_0$
by closing the integration path
in the lower half of the complex plane.
In fact,
if the contribution of the additional path
in the lower half of the complex plane can be neglected,
this procedure picks up only the contribution
from the neutrino pole which lies inside the path,
whereas the antineutrino contribution is neglected.
However,
the choice of the contour needs some caution
because the term $-(q_0)^2$ in the exponent
diverges as $q_0\to -i\infty$
and this prevents us
from using the usual half circle contour
that encircles the lower half of the complex plane.
Instead, the appropriate integration path
has the form of a rectangle
whose lower side dissects the imaginary axis
at
$q_0=-i\left[6\sigma^2_p\lambda_P\lambda_D/(\lambda_P+\lambda_D)\right] T$.
The contributions from the three sides except the real axis
are negligible.
The integrals along the sides at $\pm\infty$ vanish
since the sides are finite
in length and are damped by the $-(q_0)^2$ term
in the exponent.
The lower side gives a finite result with a exponentially damping factor
$-\left[3\sigma^2_p\lambda_P\lambda_D/(\lambda_P+\lambda_D)\right] T^2$
(easily obtained with a saddle point approximation)
which suppresses strongly
its contribution
for macroscopic time separations.
Therefore the integration over $q_0$
is dominated by the neutrino pole
which lies inside the integration contour.
The resulting amplitude is given by
\begin{equation}
{\cal A} \propto
\sum_{a} {\cal U}_{\mu a}^{*} \, {\cal U}_{ea}
\int { d\vec{q} \over (2\pi)^3 } \,
\overline{U_{D}} \,
\frac{ \gamma^0 E_a(\vec{q}) - \vec{\gamma} \cdot \vec{q} }
     { E_a(\vec{q}) } \,
V_{P} \,
\exp\left[ - i E_a(\vec{q}) T
           + i \vec{q} \cdot \vec{L}
           - S_{a}(\vec{q}) \right]\ ,
\label{E:12}
\end{equation}
where
$ E_a(\vec{q}) \equiv \sqrt{ \vec{q}^2 + m_{a}^2 } $
and
\begin{equation}
\begin{array}{rcl} \displaystyle \vphantom{\Bigg|}
S_{a}(\vec{q})
& \equiv & \displaystyle
  \frac{ \left( \vec{p}_P - \vec{q} \right)^2 }
       {12 \sigma_{p}^2 }
+ \frac{ \left[ \left( E_P - E_a(\vec{q})  \right)
              - \left( \vec{p}_P - \vec{q} \right)
	        \cdot \vec{v}_{P} \right]^2 }
       {12 \sigma_{p}^2 \lambda_P }
\\ \displaystyle \vphantom{\Bigg|}
& & \displaystyle
+ \frac{ \left( \vec{p}_D + \vec{q} \right)^2 }
       { 12 \sigma_{p}^2 }
+ \frac{ \left[ \left( E_D + E_a(\vec{q}) \right)
	      - \left( \vec{p}_D + \vec{q} \right)
	        \cdot \vec{v}_{D} \right]^2 }
       { 12 \sigma_{p}^2 \lambda_D }\ .
\end{array}
\label{E:13}
\end{equation}
Since $\sigma_p$ is small and $\lambda_P$ and $\lambda_D$ are
of order of unity,
the integral over $ d\vec{q} $
is dominated by the minimum of $ S_{a}(\vec{q}) $,
which occurs at $ \vec{q} = \vec{q}_a $, given by
\begin{eqnarray} \vphantom{\Bigg|}
\lefteqn{
  \left( \vec{v}_P-\vec{v}_a \over \lambda_P \right)
  \left[ E_P-E_a-\vec{v}_P\cdot\left(\vec{p}_P-\vec{q}_a\right) \right]
- \left(\vec{p}_P-\vec{q}_a\right)
+ }
\nonumber \\ & & \vphantom{\Bigg|}
+ \left( \vec{v}_D+\vec{v}_a \over \lambda_D \right)
  \left[ E_D+E_a-\vec{v}_D\cdot\left(\vec{p}_D+\vec{q}_a\right) \right]
- \left(\vec{p}_D+\vec{q}_a\right)
= 0\ ,
\label{E:14}
\end{eqnarray}
where
$ E_a \equiv E_a(\vec{q}_a) = \sqrt{ \vec{q}_a^2 + m_{a}^2 } $
and
$ \vec{v}_{a} \equiv \vec{q}_a / E_a $
are the velocities of the mass eigenstate neutrinos
propagating between the two interaction vertices.
A saddle point approximation of
the integral over $ d\vec{q} $ leads to
\begin{equation}
{\cal A} \propto
\sum_{a} {\cal U}_{\mu a}^{*} \, {\cal U}_{ea} \, {\cal A}_a \,
\exp\left[
- i E_a\ T
+ i \vec{q}_a \cdot \vec{L}
- S_{a}\left(\vec{q}_a\right)
- \frac{1}{2}
  \left(\vec{L}-\vec{v}_a T \right)
  \Omega_a^{-1}
  \left(\vec{L}-\vec{v}_a T \right)
    \right]
\label{E:23}
\end{equation}
where
\begin{equation}
\begin{array}{rcl} \displaystyle
{\cal A}_a
& \equiv & \displaystyle
{ 1 \over \sqrt{ {\rm Det}\left(\Omega_a\right) } } \,
\overline{U_{D}} \,
\frac{ \gamma^0 E_a - \vec{\gamma} \cdot \vec{q}_a }
     { E_a } \,
V_{P}
\\ \displaystyle
(\Omega_a)_{ij}
& \equiv & \displaystyle
  \frac{\delta_{ij}}{3\sigma_p^2}
+ \frac{ (\vec{v}_P-\vec{v}_a)_i (\vec{v}_P-\vec{v}_a)_j }
       {6\lambda_P\sigma_p^2}
+ \frac{ (\vec{v}_D+\vec{v}_a)_i (\vec{v}_D+\vec{v}_a)_j }
       {6\lambda_D\sigma_p^2}\ .
\end{array}
\label{E:omega}
\end{equation}
In Eq.(\ref{E:23}) $ \vec{V} {\rm M} \vec{V} $ denotes
$ \sum_{ij} V_i {\rm M}_{ij} V_j $ for arbitrary vector $ \vec{V} $ and
matrix M.

The amplitude (\ref{E:23}) describes the process
under consideration with the assumptions
that the wave packets of the
external particles are sharply peaked around their average momenta
and the production and detection processes are macroscopically
separated in space-time.
The amplitude contains the space-time dependent phase factor
$ \exp\left[ - i E_a\ T + i \vec{q}_a \cdot \vec{L} \right] $
which gives rise to the conventional neutrino oscillations.
The exponential damping factor
$ \exp\left[ - S_{a}\left(\vec{q}_a\right) \right] $
implements the overall energy-momentum conservation only
within an uncertainty
$ \sigma_{p} $
(if
$ \left| E_P + E_D \right|
\mbox{ \begin{picture}(12,12)\put(0,-3){$\sim$}\put(0,2){$<$}\end{picture}}
\sigma_{p} $
and
$ \left| \vec{p}_P+\vec{p}_D\right|
{\mbox{ \begin{picture}(12,12)\put(0,-3){$\sim$}\put(0,2){$<$}\end{picture}} }
\sigma_{p} $
then
$ \exp\left[ - S_{a}\left(\vec{q}_a\right) \right] \simeq 1 $).
Due to the damping factor
$ \exp\left[ - \frac{1}{2}
               \left( \vec{L} - \vec{v}_{a}\ T \right)
               \Omega_{a}^{-1}
               \left( \vec{L} - \vec{v}_{a}\ T \right)
      \right] $,
since the matrices $ \Omega_{a} $
are proportional to $ 1 / \sigma_{p}^2 \sim \sigma_{x}^2 $
(see Eq.(\ref{E:omega})),
the amplitude in Eq.(\ref{E:23}) is non-vanishing
if the velocities of the mass eigenstate neutrinos satisfy
\begin{equation} \vphantom{\Bigg|}
\left| \vec{L} - \vec{v}_{a} \, T \right|
{\mbox{ \begin{picture}(12,12)\put(0,-3){$\sim$}\put(0,2){$<$}\end{picture}} }
\sigma_{x}\ .
\label{E:29}
\end{equation}
However, if the mass difference between the mass eigenstates
$ \nu_{a} $ and $ \nu_{b} $
is such that
$ \left| \vec{v}_{a} - \vec{v}_{b} \right| \, T
{\mbox{ \begin{picture}(12,12)\put(0,-3){$\sim$}\put(0,2){$>$}\end{picture}} }
\sigma_{x} $,
then the condition (\ref{E:29})
cannot be satisfied by both mass eigenstates.
In this case,
at a given time $ T $ the amplitude has two (or more) separate peaks in space
corresponding to the two (or more) mass eigenstate neutrinos
and the experiment measures only a constant probability (in space)
for the flavor changing process under consideration.
This is due to the fact that the wave packets of the two mass eigenstates
are separated by a distance larger than their width
and, since they do not overlap,
the interference term that produces the neutrino oscillations is damped out.

\vspace*{1.5cm}
\section{ Cross Section }
\label{S:CS}

The cross section for the process Eq.(\ref{E:0}) is given,
from Eq.(\ref{E:23}), by
\begin{equation}
\begin{array}{rcl} \displaystyle
\sigma(\vec{L},T)
& \propto & \displaystyle
\int d \vec{P}
\sum_{a,b}
{\cal A}_{a} \, {\cal A}_{b}^{*} \,
{\cal U}_{\mu a}^{*} \, {\cal U}_{ea} \, {\cal U}_{\mu b} \, {\cal U}_{eb}^{*}
\,
\exp\left[
- S_{a}\left(\vec{q}_a\right)
- S_{b}\left(\vec{q}_b\right)
    \right]
\\ \displaystyle
& & \displaystyle \times
\exp\left[
- i \left( E_a - E_b \right) T
+ i \left( \vec{q}_a - \vec{q}_b \right) \cdot \vec{L}
\right]
\\ \displaystyle
& & \displaystyle \times
\exp\left[
- \frac{1}{2}
	\left( \vec{L} - \vec{v}_{a}\ T \right)
	\Omega_{a}^{-1}
	\left( \vec{L} - \vec{v}_{a}\ T \right)
- \frac{1}{2}
	\left( \vec{L} - \vec{v}_{b}\ T \right)
	\Omega_{b}^{-1}
	\left( \vec{L} - \vec{v}_{b}\ T \right)
    \right] \ ,
\end{array}
\label{E:35}
\end{equation}
where
$ \int d \vec{P} $
represents the integration over the 3-momenta and the sum over the spins
of the final particles;
one must also include appropriate average
over the 3-momenta and spins of the initial particles
which are not measured.

In a practical experimental setting,
the distance $ \vec{L} $ is usually a fixed and known quantity,
whereas the time $ T $ is not.
Therefore, the cross section
at a given distance $ \vec{L} $
is given by the time average of $ \sigma(\vec{L},T) $.
We take $ \vec{L} $ along the $ z $ direction
and integrate over time to obtain
\begin{equation}
\begin{array}{rcl} \displaystyle
\sigma(L)
& \propto & \displaystyle
\int d \vec{P}
\sum_{a,b}
{\cal A}_{a} \, {\cal A}_{b}^{*} \,
{\cal U}_{\mu a}^{*} \, {\cal U}_{ea} \, {\cal U}_{\mu b} \, {\cal U}_{eb}^{*}
\left[
	\vec{v}_a \Omega_{a}^{-1} \vec{v}_a
	       + \vec{v}_b \Omega_{b}^{-1} \vec{v}_b
\right]^{-1/2}
\\ \displaystyle
& & \displaystyle \times
\exp\left[
- S_{a}\left(\vec{q}_a\right)
- S_{b}\left(\vec{q}_b\right)
\right]
\\ \displaystyle
& & \displaystyle \times
\exp\left\{
i \left[
  \left( q_{az} - q_{bz} \right)
- \left( E_a - E_b \right)
  {\displaystyle {
 	\left( \Omega_{a}^{-1} \vec{v}_a \right)_z
	+ \left( \Omega_{b}^{-1} \vec{v}_b \right)_z
	  }\over
	\displaystyle{
	\vec{v}_a \Omega_{a}^{-1} \vec{v}_a
	       + \vec{v}_b \Omega_{b}^{-1} \vec{v}_b
	  }}
  \right] L
\right\}
\\ \displaystyle
& & \displaystyle \times
\exp\left\{
- \frac{ L^2 }{ 2 } \left[
    \left[ \Omega_{a}^{-1} \right]_{zz}
  + \left[ \Omega_{b}^{-1} \right]_{zz}
  - {\displaystyle { \left[
 	\left( \Omega_{a}^{-1} \vec{v}_a \right)_z
	+ \left( \Omega_{b}^{-1} \vec{v}_b \right)_z
	    \right]^2 } \over \displaystyle
    { 	\vec{v}_a \Omega_{a}^{-1} \vec{v}_a
	       + \vec{v}_b \Omega_{b}^{-1} \vec{v}_b }}
  \right]
\right\}
\\ \displaystyle
& & \displaystyle \times
\exp\left\{
- \frac{1}{2}\
  {\displaystyle { \left( E_a - E_b \right)^2 }
\over\displaystyle { 	\vec{v}_a \Omega_{a}^{-1} \vec{v}_a
	       + \vec{v}_b \Omega_{b}^{-1} \vec{v}_b }}
\right\}\ .
\end{array}
\label{E:pofl1}
\end{equation}

The first exponential term in Eq.(\ref{E:pofl1})
guarantees energy-momentum conservation within
the accuracy of the uncertainty principle.
The second gives rise to the neutrino oscillation in terms of the
distance L.
The third is a damping factor which
describes the coherence of the process allowing significant contributions only
from the propagation of the mass eigenstate neutrinos in the $z$ direction.
In fact,
since $ \Omega_{a} \sim \sigma_x^2 $
and $ L^2 / \sigma_x^2 \gg 1 $,
the integration over the 3-momenta of the final particles
receives its dominant contribution when this damping exponential
factor becomes maximal.
For each pair $a$ and $b$,
this maximum occurs when both
$ \vec{v}_{a} $ and $ \vec{v}_{b} $
are in the $ z $ direction.
{}From the stationary equation (\ref{E:14}),
this is realized when all
$ \vec{p}_P $,
$ \vec{v}_{P} $,
$ \vec{p}_D $ and
$ \vec{v}_{D} $
are in the $ z $ direction.
In this case the matrices
$ \Omega_{a} $ and $ \Omega_{b} $
are diagonal (see Eq.(\ref{E:omega})).
Let us denote with underlines
all the quantities evaluated at the maximum
of the damping exponential
and perform a saddle point approximation
of the integration over the angular
variables that parameterize the deviation of
$ \vec{v}_{a} $ and $ \vec{v}_{b} $
from the $ z $ direction.
As shown in Appendix \ref{A:B},
the result can be written as
\begin{equation}
\begin{array}{rcl} \displaystyle
\sigma(L)
& \propto & \displaystyle
\frac{1}{L^2}
\int d \vec{\underline{P}}
\sum_{a,b}
\underline{F}_{ab}(L)
\underline{{\cal A}}_{a} \underline{{\cal A}}_{b}^{*}
{\cal U}_{\mu a}^{*} \, {\cal U}_{ea} \, {\cal U}_{\mu b} \, {\cal U}_{eb}^{*}
\left[
  \vec{\underline{v}}_a \underline{\Omega}_{a}^{-1} \vec{\underline{v}}_a
+ \vec{\underline{v}}_b \underline{\Omega}_{b}^{-1} \vec{\underline{v}}_b
\right]^{-1/2}
\\ \displaystyle
& & \displaystyle \times
\exp\left[
- S_{a}\left(\vec{\underline{q}}_a\right)
- S_{b}\left(\vec{\underline{q}}_b\right)
\right]
\\ \displaystyle
& & \displaystyle \times
\exp\left\{
i \left[
  \left( \underline{q}_{az} - \underline{q}_{bz} \right)
- \left( \underline{E}_a - \underline{E}_b \right)
{\displaystyle
	{ \left[ \underline{\Omega}_{a}^{-1} \right]_{zz} \underline{v}_{az}
	+ \left[ \underline{\Omega}_{b}^{-1} \right]_{zz} \underline{v}_{bz} }
\over\displaystyle
	{ \left[ \underline{\Omega}_{a}^{-1} \right]_{zz} \underline{v}_{az}^2
	+ \left[ \underline{\Omega}_{b}^{-1} \right]_{zz} \underline{v}_{bz}^2 }
}
  \right] L
\right\}
\\ \displaystyle
& & \displaystyle \times
\exp\left\{
- \frac{ L^2 }{ 2 }
{\displaystyle
	{ \left( \underline{v}_{az} - \underline{v}_{bz} \right)^2 }
\over\displaystyle
	{ \left[ \underline{\Omega}_{b} \right]_{zz} \underline{v}_{az}^2
	+ \left[ \underline{\Omega}_{a} \right]_{zz} \underline{v}_{bz}^2 }
}
\right\}
\\ \displaystyle
& & \displaystyle \times
\exp\left\{
- \frac{1}{2}
{\displaystyle
	{ \left( \underline{E}_a - \underline{E}_b \right)^2 }
\over\displaystyle
	{ \left[ \underline{\Omega}_{a}^{-1} \right]_{zz} \underline{v}_{az}^2
	+ \left[ \underline{\Omega}_{b}^{-1} \right]_{zz} \underline{v}_{bz}^2 }
}
\right\} \ ,
\end{array}
\label{E:pofl}
\end{equation}
where
$ \underline{F}_{ab}(L) $
is a factor which is weakly dependent on $L$.
Since
$ \underline{F}_{ab}(L) $
becomes constant for large $L$,
as shown in Appendix \ref{A:B}, we shall neglect
$ \underline{F}_{ab}(L) $
in the following.
In Eq.(\ref{E:pofl}), $\int d \vec{\underline{P}}$
represents the remaining integrations over the momenta of the initial and
final particles.
The factor $1/L^2$ represents the geometric decrease of the neutrino flux
due to the distance of propagation, $ L$.
It is important to point out here that the second exponential
in Eq.(\ref{E:pofl}) cannot be factored out
to derive the oscillation probability as in the usual treatment of
oscillations.

Equation (\ref{E:pofl})
also contains a damping factor which
decreases exponentially with $ L^2 $
and measures the coherence
of the contributions of the different mass
eigenstate neutrinos.
The coherence length
for $ a \not= b $ is defined by
\begin{equation} \vphantom{\Bigg|}
L^{\rm coh}_{ab} \sim
\sqrt{
{
 \displaystyle
	{ \left[ \underline{\Omega}_{b} \right]_{zz} \underline{v}_{az}^2
      	+ \left[ \underline{\Omega}_{a} \right]_{zz} \underline{v}_{bz}^2 }
 \over\displaystyle
      { \left( \underline{v}_{az} - \underline{v}_{bz} \right)^2 }
}
}
\sim
\sigma_{x}
\sqrt{
	{
	 \displaystyle
		{ \underline{v}_{az}^2 + \underline{v}_{bz}^2 }
 	 \over\displaystyle
      		{ \left( \underline{v}_{az} - \underline{v}_{bz} \right)^2 }
	}
}\ ,
\label{E:48}
\end{equation}
beyond which neutrinos do not practically oscillate.
This coherence length can be very large
in the case of relativistic neutrinos,
for which
$ \left| \underline{v}_{az} - \underline{v}_{bz} \right| \ll 1 $.

The last factor in Eq.(\ref{E:pofl})
is due to the time integration
and suppresses the interference of the contributions
coming from the propagation of different mass eigenstate neutrinos
unless
$ \left| \underline{E}_a - \underline{E}_b \right|
{\mbox{ \begin{picture}(12,12)\put(0,-3){$\sim$}\put(0,2){$<$}\end{picture}} }
\sigma_{p} $,
as it should be from energy conservation in both the production and detection
interactions.

\vspace*{1.5cm}
\section{ Relativistic Limit }
\label{S:RL}

As we have emphasized, although our general result given in Eq.(\ref{E:pofl})
exhibits characteristics of neutrino oscillations,
the oscillation probability could not, in general, be factored out.
We now demonstrate that this can be done when intermediate neutrinos
are extremely relativistic.

Let us consider a process in which all
the intermediate mass eigenstate neutrinos are relativistic,
i.e. $ m_{a} \ll \underline{E}_a $.
In this case, the momentum $ \underline{q}_{az} $,
the energy $ \underline{E}_a $
and the velocity $ \underline{v}_{az} $
can be expanded as
\begin{equation}
\begin{array}{l} \displaystyle
\underline{q}_{az} = \underline{q}_{0z} + \underline{\epsilon}_{az}
\\ \displaystyle
\underline{E}_a =
  \underline{q}_{0z} + \underline{\epsilon}_{az}
+ \frac{ m_{a}^2 }{ 2 \underline{q}_{0z} }
\\ \displaystyle
\underline{v}_{az} = 1 - \frac{ m_{a}^2 }{ 2 \underline{q}_{0z}^2 }\ ,
\end{array}
\label{E:r27}
\end{equation}
where $ \underline{q}_{0z} $ is the solution of
the stationary equation (\ref{E:14})
in the $z$ direction
for $ m_{a} = 0 $ and
$ \underline{\epsilon}_{az}
  \sim m_{a}^2 / \underline{q}_{0z}
  \ll \underline{q}_{0z} $
is given by the solution of the stationary equation (\ref{E:14})
in the $z$ direction
to first order in $ m_{a}^2 / \underline{q}_{0z} $.
To lowest order in the relativistic approximation
the space-dependent part of $ \sigma(L) $
can be factorized as
$ \sigma(L) = P(L) \sigma_0 $,
where $ \sigma_0 $ is the cross section
for  massless neutrinos.
The space-dependent probability $ P(L) $
is then given by
\begin{equation}
\begin{array}{rcl} \displaystyle
P(L)
& = & \displaystyle
\sum_{a,b}
{\cal U}_{\mu a}^{*} \, {\cal U}_{ea} \, {\cal U}_{\mu b} \, {\cal U}_{eb}^{*}
\,
\exp\left\{
- i \frac{ m_{a}^2 - m_{b}^2 }{ 2 \underline{q}_{0z} } L
\right\}
\\ \displaystyle
& & \displaystyle
\phantom{ \sum_{a,b} }
\times
\exp\left\{
- \frac{ L^2 }{ 2 }
\left[ \underline{\Omega}_{0}^{-1} \right]_{zz}
\left( \frac{ m_{a}^2 - m_{b}^2 }{ 2 \underline{q}_{0z}^2 } \right)^2
- \frac{ \left( \underline{\epsilon}_{az} - \underline{\epsilon}_{bz}
              + \frac{ m_{a}^2 - m_{b}^2 }{ 2 \underline{q}_{0z} } \right)^2 }
       { 4 \left[ \underline{\Omega}_{0}^{-1} \right]_{zz} }
\right\}
\end{array}
\label{E:pofltwo}
\end{equation}
The first line of Eq.(\ref{E:pofltwo}) gives
the usual oscillation probability for
relativistic neutrinos
(which can be obtained from a quantum mechanical treatment of the
neutrino oscillations
\cite{BilenkyPontecorvo78,BilenkyPetcov87,Kayser81,Frampton82,Boehm87}).
The second line of Eq.(\ref{E:pofltwo}) contains an exponent which
decreases quadratically with the distance $L$
and measures the coherence of the
contributions due to the wave packets
of the different mass eigenstate neutrinos.
Since
$ \left[ \underline{\Omega}_{0}^{-1} \right]_{zz}^{-1} =
  \left[ \underline{\Omega}_{0} \right]_{zz} \sim \sigma_x^2 $,
the coherence length becomes
\begin{equation} \vphantom{\Bigg|}
L^{\rm coh}_{ab} \sim
\sigma_x
{
 \displaystyle
	{ 2 \underline{q}_{0z}^2 }
 \over\displaystyle
      	{ m_{a}^2 - m_{b}^2 }
}
\ .
\label{E:r35}
\end{equation}
The length, $ L^{\rm coh}_{ab} $,
is the coherence length for the neutrino oscillations,
i.e.
the two mass eigenstate neutrinos $ \nu_{a} $ and $ \nu_{b} $
contribute coherently to the flavor changing process
only when $ L \ll L^{\rm coh}_{ab} $,
in which case the probability oscillates
as a function of the distance $ L $.
The coherence length given in Eq.(\ref{E:r35}) is the same
as that obtained by physical intuitions
in Ref.\cite{BilenkyPontecorvo78,Kayser81,Nussinov76}
and from a quantum mechanical wave-packet treatment
in Ref.\cite{GKL9116}.
If the distance $ L $ is much smaller than the coherence length
$ L^{\rm coh}_{ab} $,
the damping factor in the probability (\ref{E:pofltwo})
becomes approximately unity and
hence one obtains the usual oscillation probability.
On the other hand,
for distances $ L \gg L^{\rm coh}_{ab} $
the two mass eigenstate neutrinos $ \nu_{a} $ and $ \nu_{b} $
contribute incoherently to the flavor changing process.

{}From Eq.(\ref{E:pofltwo}),
the well-known oscillation wavelength $ L^{\rm osc}_{ab} $ is
\begin{equation} \vphantom{\Bigg|}
L^{\rm osc}_{ab} =
2 \pi
\frac{ 2 \underline{q}_{0z} }{ \left| m_{a}^2 - m_{b}^2 \right| }
\label{E:r39}
\end{equation}
so that Eq.(\ref{E:r35}) can be written as
\begin{equation} \vphantom{\Bigg|}
L^{\rm coh}_{ab} \sim
\frac{ \underline{q}_{0z} }{ \sigma_{p} }
L^{\rm osc}_{ab}\ .
\label{E:r40}
\end{equation}
Hence, the maximum number of observable oscillations before the decoherence of
the wave packets of the different mass eigenstate neutrinos is given by
\cite{Frampton82,Boehm87,Nussinov76}
\begin{equation} \vphantom{\Bigg|}
N_{\rm osc} =
\frac{ L^{\rm coh}_{ab} }
     { L^{\rm osc}_{ab} }
\sim
\frac{ \underline{q}_{0z} }{ \sigma_{p} }\ .
\label{E:r41}
\end{equation}
Notice that $ N_{\rm osc} $ is independent
of the neutrino mass eigenvalues $ m_{a} $.
Since we have assumed that $ \sigma_{p} $ is much smaller than
the energies of the initial and final particles involved,
$ \sigma_{p} \ll \underline{q}_{0z} $
and
$ N_{\rm osc} \gg 1 $.

Finally, in the last exponential factor of the probability (\ref{E:pofltwo})
we have retained a space-independent damping factor.
Since
\begin{equation} \vphantom{\Bigg|}
  \epsilon_{az} - \epsilon_{bz}
+ \frac{ m_{a}^2 - m_{b}^2 }{ 2 \underline{q}_{0z} } \sim
  \frac{ m_{a}^2 - m_{b}^2 }{ \underline{q}_{0z} } \sim
\frac{ 1 }{ L^{\rm osc}_{ab} }
\label{E:r51}
\end{equation}
the probability (\ref{E:pofltwo}) for $ a \not= b $
does not vanish only when
$  L^{\rm osc}_{ab}
{\mbox{ \begin{picture}(12,12)\put(0,-3){$\sim$}\put(0,2){$>$}\end{picture}} }
\sigma_{x} $.
This result is due to the time integration:
if the neutrino wave packets are larger than the oscillation length,
the interference terms are washed out.

\vspace*{1.5cm}
\section{ Conclusions }

In order to illustrate
the quantum field theoretical treatment
of neutrino oscillations without introducing the concept of weak eigenstates,
we have discussed
a specific flavor changing process (see Eq.(\ref{E:0}))
in which the external particles are described by wave packets
and the mass eigenstate neutrinos
propagate between the production and detection
interactions which are macroscopically localized in space-time.
We have calculated the time-averaged cross section
which is the measurable quantity
in the usual experimental setting
where the  distance
between the production and detection interactions
is known but the time separation is not measured.
We have pointed out that in general,
it is not possible to factor out of the cross section
a space-dependent oscillation probability
because the dynamics of the production and detection interactions
is not the same for the different mass eigenstate neutrinos.
However, we have shown that
in the extremely relativistic limit
the usual quantum mechanical oscillation probability
can be factored out of the cross section.
\vskip1.5cm

\noindent
{ \Large \bf Acknowledgement}

\noindent
We wish to thank G. Feldman for useful discussions.
Two of the authors (CWK and UWL) wish to acknowledge the support from
the Center for Theoretical Physics, Seoul National University.
This work was supported in part by the National Science Foundation.

\appendix
\newpage

\vspace*{1.5cm}
\section{Wave Packet}
\label{A:A}

A gaussian wave packet in the momentum space is given by
\begin{equation} \vphantom{\Bigg|}
\psi(\vec{p};\vec{X},T,\left\langle{\vec{p}}\right\rangle) =
\left[ \sqrt{ 2 \pi } \sigma_{p} \right]^{-3/2}
\exp\left[ - \frac{ \left( \vec{p}
	- \left\langle{\vec{p}}\right\rangle \right)^2 }
                  { 4 \sigma_{p}^2 }
           - i \vec{p} \cdot \vec{X}
           + i E(\vec{p}) T \right]\ ,
\label{E:A.1}
\end{equation}
where $ E(\vec{p}) \equiv \sqrt{ \vec{p}^2 + m^2 } $
and $\sigma_{p}$ is the width of the wave packet
for simplicity assumed to be the same along the three directions.

In the coordinate space, we have
\begin{equation} \vphantom{\Bigg|}
\psi(\vec{x},t;\vec{X},T,\left\langle{\vec{p}}\right\rangle) =
\int { d\vec{p} \over (2\pi)^{3/2} }
\ \psi(\vec{p};\vec{X},T,\left\langle{\vec{p}}\right\rangle)
\ {\rm e}^{ i \vec{p} \cdot \vec{x} - i E(\vec{p}) t }\ .
\label{E:A.2}
\end{equation}
Since the gaussian wave packet in the momentum space
is peaked around the average momentum $\left\langle{\vec{p}}\right\rangle$,
neglecting the spreading of the wave packet,
one can approximate
\begin{eqnarray} &\vphantom{\Bigg|}&
E(\vec{p}) \simeq \left\langle{E}\right\rangle + \vec{v} \left( \vec{p}
	- \left\langle{\vec{p}}\right\rangle \right);
\nonumber \\ &\vphantom{\Bigg|}&
\left\langle{E}\right\rangle \equiv E
	\left(\left\langle{\vec{p}}\right\rangle\right)
= \sqrt{ {\left\langle{\vec{p}}\right\rangle}^2 + m^2 }
\label{E:A.4} \\ &\vphantom{\Bigg|}&
\vec{v} \equiv
\left.
{\partial {E}\over \partial {\vec{p}} }
	\right|_{\vec{p}=\left\langle{\vec{p}}\right\rangle} =
\frac{ \left\langle{\vec{p}}\right\rangle }{
	\left\langle{E}\right\rangle }\ \ \ \ .
\nonumber
\end{eqnarray}
Hence the wave packet in the coordinate space is found to be
\begin{equation} \vphantom{\Bigg|} \hskip-0.25in
\psi(\vec{x},t;\vec{X},T,\left\langle{\vec{p}}\right\rangle) \simeq
\left[ \sqrt{ 2 \pi } \sigma_{x} \right]^{-3/2}
\exp\left[
i \left\langle{\vec{p}}\right\rangle \cdot \left( \vec{x} - \vec{X} \right)
         - i \left\langle{E}\right\rangle \left( t - T \right)
         - \frac{ \left[ \left( \vec{x} - \vec{X} \right)
                - \vec{v} \left( t - T \right) \right]^2 }
                { 4 \sigma_{x}^2 } \right]\ .
\label{E:A.5}
\end{equation}
At time $ t = T $ the wave packet is peaked at $ \vec{x} = \vec{X} $
with a width $ \sigma_{x} $ given by
\begin{equation} \vphantom{\Bigg|}
\sigma_{x}\ \sigma_{p} = \frac{1}{2}
\label{E:A.6}
\end{equation}
in agreement with the uncertainty principle.
The wave packet propagates in space with group velocity $ \vec{v} $.

\vspace{1.5cm}
\section{ Saddle Point Approximation}
\label{A:B}

The 2-dimensional integration over the angular
variables
$ \vec{\xi} = (\xi_1,\xi_2) $
that parameterize the deviation of
$ \vec{v}_{a} $ and $ \vec{v}_{b} $
from the $ z $ direction in Eq.(\ref{E:pofl1})
is of the type
\begin{equation}
I = \int d \vec{\xi}
\exp\left[ i A(\vec{\xi}) L - B(\vec{\xi}) L^2 - C(\vec{\xi}) \right]\ .
\label{E:C.1}
\end{equation}
For large $L$
the integral gets its dominant contribution
from the minimum of $B(\vec{\xi})$,
which occurs for $\vec{\xi}=\vec{\underline{\xi}}$.
We expand all the terms around this minimum:
\begin{equation}
\begin{array}{rcl} \displaystyle
A(\vec{\xi})
& = & \displaystyle
  \underline{A}
+ \sum_{i=1,2}
  \underline{A}'_{i} \left( \xi^{i} - \underline{\xi}^{i} \right)
\\ \displaystyle
B(\vec{\xi})
& = & \displaystyle
  \underline{B}
+ {1\over2} \sum_{i,j=1,2}
  \left( \xi^{i} - \underline{\xi}^{i} \right)
  \underline{B}''_{ij}
  \left( \xi^{j} - \underline{\xi}^{j} \right)
\\ \displaystyle
C(\vec{\xi})
& = & \displaystyle
  \underline{C}
+ \sum_{i=1,2}
  \underline{C}'_{i} \left( \xi^{i} - \underline{\xi}^{i} \right)\ .
\end{array}
\end{equation}
After a change of variable
$ \left( \xi^{i} - \underline{\xi}^{i} \right) \to \xi^{i} $,
the integral in Eq.(\ref{E:C.1}) can be written in gaussian form
\begin{equation}
\begin{array}{rcl} \displaystyle
I
& = & \displaystyle
\exp\left[ i \underline{A} L - \underline{B} L^2 - \underline{C} \right]
\\ \displaystyle
& & \displaystyle \times
\exp\left[
{1\over2}
\left( i \underline{A}'_{i} - \frac{1}{L} \underline{C}'_{i} \right)
\left[ \underline{B}''^{-1} \right]^{ij}
\left( i \underline{A}'_{j} - \frac{1}{L} \underline{C}'_{j} \right)
\right]
\\ \displaystyle
& & \displaystyle \times
\int d \vec{\xi}
\exp\left\{ \scriptstyle
- \frac{L^2}{2}
\left[ \xi^{i} - \frac{1}{L}
\left( i \underline{A}'_{k} - \frac{1}{L} \underline{C}'_{k} \right)
\left[ \underline{B}''^{-1} \right]^{ki}
\right]
\underline{B}''_{ij}
\left[ \xi^{j} - \frac{1}{L}
\left( i \underline{A}'_{l} - \frac{1}{L} \underline{C}'_{l} \right)
\left[ \underline{B}''^{-1} \right]^{lj}
\right]
\right\}\ .
\end{array}
\end{equation}
The final result is
\begin{equation}
I =
\frac{2\pi}{L^2}
\frac{F(L)}{ \sqrt{ {\rm Det}\left( { \underline{B}''_{ij} }\right) } }
\exp\left[ i \underline{A} L - \underline{B} L^2 - \underline{C} \right]
\end{equation}
with
\begin{equation}
F(L) =
\exp\left[
{1\over2}
\left( i \underline{A}'_{i} - \frac{1}{L} \underline{C}'_{i} \right)
\left[ \underline{B}''^{-1} \right]^{ij}
\left( i \underline{A}'_{j} - \frac{1}{L} \underline{C}'_{j} \right)
\right]\ .
\label{E:C5}
\end{equation}
{}From Eq.(\ref{E:C5}) it is clear that
the space dependence of $F(L)$ is negligible for large $L$.


\end{document}